\documentclass{PHYEAUTH}
\usepackage{graphicx}
\usepackage{amsmath}
\usepackage{amssymb}

\begin{document}

\begin{frontmatter}

\title{Charge and spin manipulation in a few-electron double dot}

\author[address1]{J.~R.~Petta\thanksref{thank1}},
\author[address1]{A.~C.~Johnson},
\author[address1]{J.~M.~Taylor},
\author[address2]{A.~Yacoby},
\author[address1]{M.~D.~Lukin},
\author[address1]{C.~M.~Marcus},
\author[address3]{M.~P.~Hanson},
and \author[address3]{A.~C.~Gossard},

\address[address1]{Department of Physics, Harvard University, Cambridge,
MA  02138}

\address[address2]{Department of Condensed Matter Physics,
Weizmann Institute  of Science, Rehovot 76100 Israel}

\address[address3]{Materials Department, University of California, Santa
Barbara, CA 93106}

\thanks[thank1]{Corresponding author. E-mail:
petta@deas.harvard.edu}

\begin{abstract}
We demonstrate high speed manipulation of a few-electron double
quantum dot. In the one-electron regime, the double dot forms a
charge qubit. Microwaves are used drive transitions between the
(1,0) and (0,1) charge states of the double dot. A local quantum
point contact charge detector measures the photon-induced change
in occupancy of the charge states. Charge detection is used to
measure $T_1$$\sim$16 ns and also provides a lower bound estimate
for $T{_2^*}$ of 400 ps for the charge qubit. In the two-electron
regime we use pulsed-gate techniques to measure the
singlet-triplet relaxation time for nearly-degenerate spin states.
These experiments demonstrate that the hyperfine interaction leads
to fast spin relaxation at low magnetic fields. Finally, we
discuss how two-electron spin states can be used to form a logical
spin qubit.
\end{abstract}

\end{frontmatter}

Semiconducting quantum dots can be used to confine single
electrons in an electrically controllable potential
\cite{Ciorga_PRB_2000}. Coupled quantum dots, containing a single
electron, create a tunable two-level system for the manipulation
of single charges \cite{Hayashi_PRL_2003,Petta_PRL_2004}. By a
similar approach, when two electrons are confined to a double dot
the relaxation and dephasing of singlet and triplet spin states
can be studied
\cite{Ono_Science_2002,Petta_condmat,Johnson_Nature_2005}.
Recently, we have demonstrated coherent control of two-electron
spin states by using high speed pulsed gate techniques
\cite{Petta_Science}. In this paper, we review recent experiments
performed by our group on few-electron quantum dots that
demonstrate quantum control of just one or two electrons
\cite{Petta_PRL_2004,Petta_condmat,Johnson_Nature_2005,Petta_Science,Johnson_condmat}.

Samples are fabricated from a GaAs/Al$_{0.3}$Ga$_{0.7}$As
heterostructure grown by molecular beam epitaxy (Fig.\ 1(a)).
Electron beam lithography and liftoff techniques are used to
fabricate Ti/Au gates, which deplete the two-dimensional electron
gas with electron density 2$\times$10$^{11}$cm$^{-2}$ and mobility
2$\times$10$^5$cm$^2$/V$\cdot$s. When the gates are appropriately
biased a double well potential is formed. The electron number in
the left(right) dot is varied by tuning $V_L$($V_R$). Interdot
tunnel coupling is tuned by changing the voltage $V_T$. Standard
lock-in techniques are used to measure the double dot conductance,
$G_{D}$, and the quantum point contact (QPC) conductances,
$G_{S1(S2)}$. The sample is cooled to base temperature in a
dilution refrigerator with an electron temperature, $T_e$$\sim$135
mK, as determined from Coulomb blockade peak widths. Depending on
the experimental arrangement continuous-wave (cw) microwaves are
applied to gate A, or high speed pulses are applied to gates L and
R using bias tees \cite{Anritsu} that are thermally anchored to
the mixing chamber.

\begin{figure}[t]
\vspace{2.0 cm}
\begin{center}\leavevmode
\includegraphics[width=1\linewidth]{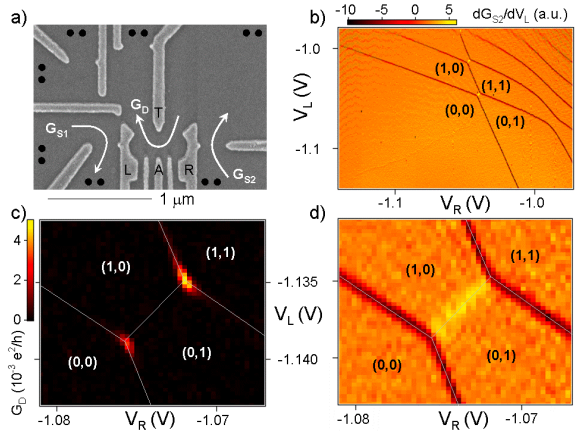}
\vspace{-2.3 cm} \caption{Transport and charge sensing of a
few-electron double dot. (a) Scanning electron microscope (SEM)
image of a double dot device similar to the one used in these
experiments. The double dot is flanked by two QPC charge
detectors. Gates L(R) primarily change the number of electrons in
the left(right) dot. Interdot tunnel coupling can easily be tuned
by adjusting the voltage on gate T. $\bullet$ denotes an Ohmic
contact. The double dot conductance, $G{_D}$, and the QPC
conductances, $G_{S1}$ and $G_{S2}$, are measured using standard
lock-in techniques. (b) Large scale plot of $dG_{S2}$/$dV_{L}$ as
a function of $V_{R}$ and $V_{L}$. Charge states are labelled
$($$M$,$N$$)$, where $M$$($$N$$)$ is the time averaged number of
electrons on the left (right) dot. $G{_D}$, in (c), and
$dG_{S2}$/$dV_{L}$, in (d), as a function of $V_{R}$ and $V_{L}$
near the (1,0) to (0,1) transition. The gates have been slightly
adjusted in (c--d) relative to (b) to allow simultaneous transport
and sensing. Identical color-scales are used in (b) and (d).}
\label{figurename}\end{center}\end{figure}

\section{Isolating single charges}
Control of the double dot using dc gate voltages is demonstrated
in Fig.\ 1 (b--d). Figure 1(b) shows $dG_{S2}$/$dV_L$ (numerically
differentiated) as a function of $V_R$ and $V_L$. When an electron
enters or leaves the double dot, or moves from one dot to the
other, the QPC conductance changes. Gate voltage derivatives of
$G_{S1}$ and $G_{S2}$ clearly show these changes and map out the
double dot charge stability diagram
\cite{Field_PRL_1993,Elzerman_PRB_2003}. The nearly horizontal
lines are due to charge transitions in the left dot, while the
nearly vertical lines correspond to charge transitions in the
right dot. For very negative values of $V_L$ and $V_R$ (see the
lower left corner of the charge stability diagram) charge
transitions no longer occur, indicating that the double dot is
completely empty, denoted (0,0). Transport through the double dot
can be correlated with simultaneous charge sensing measurements.
Figure 1(c) shows a color scale plot of $G_D$ near the (1,0) to
(0,1) charge transition. A charge stability diagram,
simultaneously acquired, is shown in Fig.\ 1(d). Near the (1,0) to
(0,1) charge transition the system behaves as an effective
two-level system. Crossing this transition by making $V_L$ more
positive transfers a single electron from the right dot to the
left dot.

\section{Microwave manipulation of a single charge}
Near the (0,1) to (1,0) interdot transition, the double dot forms
a two-level charge system that can be characterized by the
detuning parameter, $\epsilon$, and the tunnel coupling, $t$ (see
inset of Fig. 2(b)) \cite{Van_der_Wiel_RMP_2003}. We have used
microwave spectroscopy to characterize this two-level system
\cite{Petta_PRL_2004}. Microwaves drive transitions in the double
dot when the photon frequency is equal to the energy separation
between the (1,0) and (0,1) charge states
\cite{Stafford_PRL_1996,Stoof_PRB_1996,Brune_PhysicaE_1997}. This
microwave-induced charge state repopulation can be directly
measured using the QPC charge sensors
\cite{Van_der_Wal_Science_2000,Lehnert_PRL_2003}. The black curve
in Fig.\ 2(a) shows the measured charge on the left dot, $M$, as a
function of $\epsilon$, in the absence of microwave excitation. As
expected, increasing $\epsilon$ transfers a single charge from the
left dot to the right dot. Application of microwaves to gate A
results in resonant peaks in $M$ vs.\ $\epsilon$ that move to
larger $|\epsilon|$ with increasing frequency. This resonant peak
corresponds to a single photon process that drives an electron
from the (1,0) ground state (for negative $\epsilon$) into the
(0,1) excited state, or vice versa.

The frequency dependence of the resonance condition can be used to
map out the energetics of the charge two-level system. Detailed
measurements of the resonant peak position as a function of
microwave frequency, $f$, are used to extract $t$ for various
$V_T$ (see Fig.\ 2(b)) \cite{Oosterkamp_Nature_1998}. At high
frequencies the peak positions move linearly with $f$. For small
frequencies, probing the region near the (0,1)-(1,0) charge
transition, the interdot tunnel coupling modifies the linear
dependence. Changing the interdot tunnel coupling modifies the
frequency dependence of the resonant peak position. For each value
of $V_T$, the experimental data have been fit using $\alpha
\epsilon$= $\sqrt{(hf)^2-(2t)^2}$, where $\alpha$ is the lever
arm. $\alpha$ and $t$ were used as free parameters for each curve.
The lever arm $\alpha$ changes by $\sim$20$\%$ over the range of
$V_T$ used in Fig.\ 2. The experimental data are well fit by
theory and show that the tunnel coupling varies by roughly a
factor of 6 when $V_T$ is changed by 70 mV. Measurements of $t$
from microwave spectroscopy are consistent with values obtained by
measuring the width of the interdot charge transition using charge
sensing \cite{DiCarlo_PRL_2004,Petta_PRL_2004}.

\begin{figure}[t]
\vspace{1.3 cm}
\begin{center}\leavevmode
\includegraphics[width=1\linewidth]{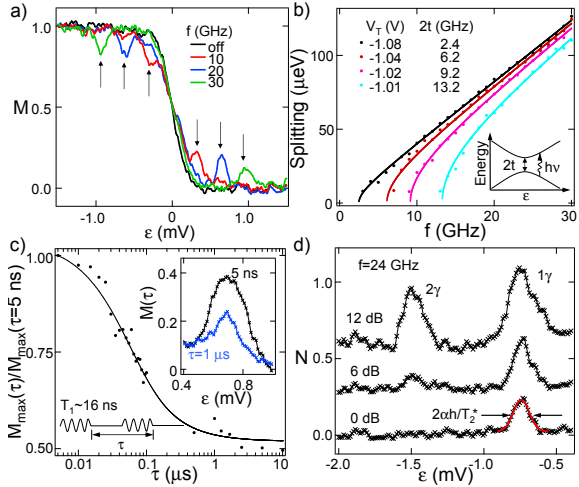}
\vspace{-1.7 cm}\caption{Microwave spectroscopy of a one-electron
double dot. (a) Charge occupancy of the left dot, $M$, as a
function of $\epsilon$ for several microwave frequencies. (b)
One-half of the resonance peak splitting as a function of $f$ for
several values of $V_T$. Solid lines are best fits to the
experimental data using the theory outlined in the text. Inset:
two-level system energy level diagram. (c) Amplitude of the
resonance, expressed as $M_{max}$($\tau$)/$M_{max}$($\tau$=5 ns),
as a function of chopped cw period, $\tau$, with $f$=19 GHz.
Theory gives a best fit $T_1$=16 ns (solid line, see text). Inset:
Single photon peak shown in a plot of $M$ as a function of
$\epsilon$ for $\tau$=5 ns and 1 $\mu$s. (d) Power dependence of
the resonance for $f$=24 GHz. Widths are used to extract the
ensemble-averaged charge dephasing time $T_{2}^{*}$. At higher
microwave powers multiple photon processes occur. Curves are
offset by 0.3 for clarity.}
\label{figurename}\end{center}\end{figure}

Charge relaxation and decoherence times can be extracted by
analyzing the resonant response of the two-level system, as used
in the analysis of the Cooper pair box \cite{Lehnert_PRL_2003}.
The charge relaxation time $T_1$ is determined by measuring the
resonance peak height as microwaves are chopped at varying
periods, $\tau$, with a 50\% duty cycle \cite{Marki}. The system
response is modelled with a saturated signal while microwaves are
present, followed by an exponential decay with a characteristic
time scale $T_1$ when the microwaves are turned off. Calculating
the time averaged occupation, we expect:
\begin{equation}
\frac{M_{max}(\tau)}{M_{max}(0)}=\frac{1}{2}+\frac{T_1(1-e^{-\tau/(2
T_1)})}{\tau}
\end{equation}
With long periods ($\tau$$\gg$$T_1$), the exponential tail due to
the finite relaxation time is an insignificant part of the duty
cycle, and the charge detector measures the time average of the
on/off signal, giving a resonant feature with half the height
found in the limit $\tau$$\rightarrow$0. When the period is very
short, such that $\tau$$\ll$$T_1$, the charge has little time to
relax, and the charge detector response is close to saturation
(saturation is defined as $M_{max}$=0.5 on resonance). In the
intermediate regime where $\tau$$\sim$$T_1$, the QPC signal is
strongly dependent on $\tau$. We present data for $\tau\geq$5 ns
to avoid artifacts due to the finite rise time of the mixer
circuit. In Fig. 2(c), we plot
$M_{max}$($\tau$)/$M_{max}$($\tau$=5 ns) as a function of $\tau$.
Agreement between experiment and theory is good and gives a best
fit $T_1$=16 ns.

The resonance peak width gives a direct measure of the
inhomogeneous charge decoherence time, $T_{2}^{*}$
\cite{Lehnert_PRL_2003,Abragam}. In Fig.\ 2(d) we plot $N$ as a
function of $\epsilon$ for several microwave powers. At low power,
only the single-photon (1$\gamma$) peak is visible. As the power
is increased the 1$\gamma$ peak approaches saturation and a
two-photon peak develops \cite{population_inversion}. A fit to the
low power 1$\gamma$ peak using a Gaussian function is shown in red
in Fig.\ 2(d). The best fit half-width of 0.077 mV corresponds to
an energy of 10.2 $\mu$eV when taking into account the lever arm.
Converting this into a time results in a lower bound
$T_{2}^{*}$=400 ps. This measurement of $T_{2}^{*}$ gives a
worst-case estimate since charge fluctuations will broaden the
resonant feature, resulting in a shorter $T_{2}^{*}$ value.

Our measurements of $T_{1}$ and $T_{2}^{*}$ using charge sensing
can be compared with other recent experiments
\cite{Hayashi_PRL_2003,Fujisawa_Nature_2002}. In a pulsed-gate
experiment, Fujisawa \textit{et al.} \cite{Fujisawa_Nature_2002}
have measured the energy relaxation time in a vertical quantum
dot. From a measurement of the transient current as a function of
pulse time they extract $T_1$=10 ns, which is limited by
spontaneous emission of a phonon. Direct observation of coherent
charge oscillations has been reported by Hayashi \textit{et al.}\
\cite{Hayashi_PRL_2003}. From the decay envelope of the Rabi
oscillations Hayashi \textit{et al.} extract a $T_{2}$ time of
$\sim$1 ns, which serves as an upper bound estimate for $T_{2}^*$.
The $T_{1}$ and $T_{2}^{*}$ values that we obtain from charge
sensing are in good agreement with the results of these previous
experiments.

\section{Triplet-singlet spin relaxation}
Spin physics can be studied in the one-electron regime at high
fields, where the spin-up and spin-down states are separated by
the Zeeman splitting, or in the two-electron regime with singlet
and triplet spin states. We focus on the two-electron regime,
where differences in the singlet-triplet splittings in the (1,1)
and (0,2) charge states can be put to use for spin state readout
and initialization. We show that singlet-triplet relaxation times
can be measured by implementing a charge pump experiment in the
two-electron regime. This measurement technique can be used to
measure the singlet-triplet relaxation time, $\tau_{ST}$, for
nearly degenerate singlet and triplet states, a regime in which
hyperfine mediated relaxation process are expected to be
important.

In the two-electron regime, charge transport in a double dot shows
a striking asymmetry in bias voltage due to spin selection rules
(Pauli  blocking) \cite{Ono_Science_2002,Johnson_condmat}. The
asymmetry in charge transport is due to the large difference in
the singlet-triplet splittings for the (1,1) and (0,2) charge
states. In the weakly-coupled (1,1) charge configuration the
singlet and triplet states are nearly degenerate. However, two
tightly confined electrons in the (0,2) charge state result in a
singlet-triplet splitting $J$$\sim$400 $\mu$eV. At forward bias,
transitions  from the (0,2) singlet state, (0,2)$_\textrm{S}$, to
the (1,1) singlet state, (1,1)$_\textrm{S}$, are allowed. However,
reverse bias (1,1) to (0,2) charge transitions are blocked if the
(1,1) state forms a triplet (1,1)$_\textrm{T}$ because the
(0,2)$_\textrm{T}$  state resides outside the transport window due
to the large singlet-triplet splitting in (0,2). This asymmetry
results in current rectification, which can be used for
spin-to-charge conversion and spin state readout.

Charge transitions are driven by applying pulses to gates L and R.
Experimental details concerning pulse calibration have been
previously published \cite{Petta_condmat}. In double dots, charge
can be pumped by pulsing gates around a triple point, e.g.
(0,1)$\rightarrow$(1,1)$\rightarrow$(0,2)$\rightarrow$(0,1). Our
spin relaxation measurement technique relies on the fact that
(1,1)$_T$ to (0,2)$_S$ transitions are spin blocked. Measuring
this charge transition probability as a function of time using
charge sensing allows a measurement of the spin relaxation time.
We demonstrate that the observed time dependence of the charge
sensing signal is due to spin blocked transitions.

\begin{figure}[t]
\vspace{1.3 cm}
\begin{center}\leavevmode
\includegraphics[width=1\linewidth]{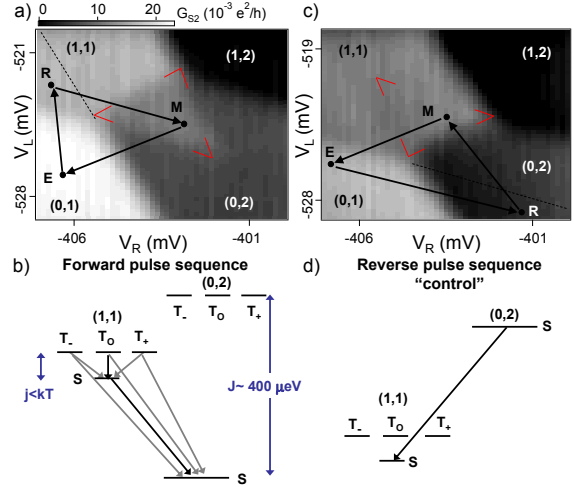}
\vspace{-1.8 cm} \caption{Pulse gate techniques for measuring the
singlet-triplet relaxation time. (a) $G_{S2}$, as a function of
$V_L$ and $V_R$, measured while applying the forward pulse
sequence
(0,1)$\rightarrow$(1,1)$\rightarrow$(0,2)$\rightarrow$(0,1). The
pulse period $\tau$=10 $\mu$s and the perpendicular field
$B_{\perp}$=100 mT. We observe a charge sensing signal in the
pulse triangle (bounded by the red lines) that takes on a value
between the raw (1,1) and (0,2) signal levels. This signal is
indicative of spin-blocked charge transitions. Outside of the
pulse triangle, transitions through the (1,2) and (0,1) charge
states are possible and relax the spin blockade. (b) Energy level
diagram illustrating the possible transitions from (1,1) to (0,2).
Fast transitions are indicated with black arrows, spin blocked
transitions with grey arrows. (c) $G_{S2}$ as a function of $V_L$
and $V_R$ while applying the reverse ``control" pulse
sequence,(0,1)$\rightarrow$(0,2)$\rightarrow$(1,1)$\rightarrow$(0,1).
The pulse period $\tau$=10 $\mu$s and the perpendicular field
$B_{\perp}$=100 mT. The (0,2)$_\textrm{S}$ to (1,1)$_\textrm{S}$
transition is not spin blocked and as a result, there is no
detectable pulse signal in the pulse triangle (bounded by the red
lines). (d) Level diagram illustrating the (0,2) to (1,1)
transition. A best-fit plane has been subtracted from the data in
(a),(c) to remove signal from direct gate to QPC coupling.}
\label{figurename}\end{center}\end{figure}

The pulse sequence used to measure $\tau_{ST}$ is shown in [Fig.\
3(a)]. The gates are held at point E for 10\% of the period,
emptying the second electron from the double dot, leaving the
(0,1) charge state. A pulse shifts the gates to point R (reset
point) for the next 10\% of the period. This initializes the
system into the (1,1) configuration. Since the singlet and triplet
states are nearly degenerate in (1,1) we expect to load into
(1,1)$_\textrm{S}$ or any  of the three (1,1)$_\textrm{T}$ states
with equal probability. For the final 80\% of the period, the
gates are held at the measurement point M where (0,2)$_\textrm{S}$
is the ground state. The energetics of the spin states at the
measurement point are shown in Fig.\ 3(b). A (1,1)$_\textrm{S}$
state prepared in the R step will tunnel to (0,2)$_\textrm{S}$ on
a timescale set by the interdot tunneling rate,
$\Gamma(\epsilon)$. The m$_\textrm{s}$=0 (1,1) triplet state,
(1,1)$_\textrm{T0}$, will dephase into (1,1)$_\textrm{S}$ on a
timescale set by $T_2$ (expected to be $\leq$100 ns
\cite{Kikkawa_PRL_98,Khaetskii_PRL_2002,Merkulov_PRB_2002})
followed by a direct transition to (0,2)$_\textrm{S}$. Roughly
half of the time the R step will load the m$_\textrm{s}$=1 (1,1)
triplet state, (1,1)$_\textrm{T+}$, or the m$_\textrm{s}$=-1 (1,1)
triplet state, (1,1)$_\textrm{T-}$. Since (0,2)$_\textrm{T}$ is
inaccessible a transition from (1,1)$_\textrm{T+}$ or
(1,1)$_\textrm{T-}$ to (0,2) requires a spin flip and will be
blocked for times shorter than the singlet-triplet relaxation time
$\tau_{ST}$.

Figure 3(a) shows the time-averaged charge sensor conductance
$G_{S2}$, measured as a function of the dc gate voltages $V_L$ and
$V_R$, while the forward pulse sequence is repeated. The
conductance $G_{S2}$ maps out the ground state population at point
M since 80\% of the duty cycle is spent there. The plateaus in
$G_{S2}$ at $ \sim$0.0, 6.0, 16, and 23 $\times$10$^{-3}$e$^2$/h
indicate full population of the (1,2), (0,2), (1,1), and (0,1)
charge states respectively. The pulse data differs from ground
state data only when point M resides in the triangle defined by
the (1,1) to (0,2) ground state transition and the extensions of
the (1,1) to (0,1) and (1,1) to (1,2) ground state transitions
(bounded by the red marks in Fig.\ 3(a)). Inside of the ``pulse
triangle" transitions from (1,1) to (0,2) may be spin-blocked and
the charge sensor registers a conductance intermediate between the
(1,1) and (0,2) plateaus. Outside of the pulse triangle it is
possible to access (0,1) or (1,2), which relaxes the spin
blockade. Figure 3(a) shows a signal of
~11$\times$10$^{-3}$e$^2$/h in the pulse triangle for $\tau$=10
$\mu$s, indicating that approximately 50\% of the time the dots
remain in (1,1) even though (0,2) is the ground state.  This is
direct evidence of spin-blocked (1,1) to (0,2) transitions.

To check that the pulse signal is due to spin-blocked transitions
and not just a slow interdot tunnel rate we compare the forward
pulse sequence with a reverse pulse sequence that does not involve
spin selective transitions
[(0,1)$\rightarrow$(0,2)$\rightarrow$(1,1)$\rightarrow$(0,1)]. In
the reverse pulse sequence the reset position R occurs in (0,2)
where only the singlet state is accessible, and M occurs in (1,1).
Now tunneling from R to M should always proceed on a time scale
set by the interdot tunnel coupling, since the (0,2)$_\textrm{S}$
to (1,1)$_\textrm{S}$ transition is not spin blocked. No signal is
seen in the pulse triangle for this reversed ``control" sequence
(Fig.\ 3(c)).

\begin{figure}[t]
\vspace{2 cm}
\begin{center}\leavevmode
\includegraphics[width=1\linewidth]{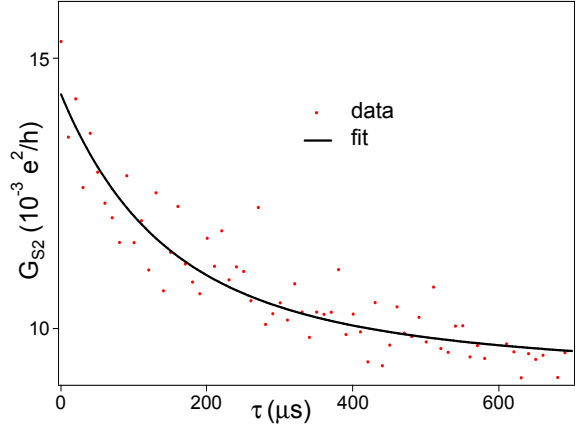}
\vspace{-2.5 cm} \caption{Time dependence of the spin-blocked
signal. $G_{S2}$ measured as a function of $\tau$. In the (0,2)
charge state $G_{S2}$=10$\times$10$^{-3}$e$^2$/h, while
$G_{S2}$=20$\times$10$^{-3}$e$^2$/h in the (1,1) charge state. For
short $\tau$, the (1,1) to (0,2) transition is blocked
approximately half of the time, resulting in a pulse signal of
15$\times$10$^{-3}$e$^2$/h in the (0,2) pulse triangle. The (1,1)
to (0,2) transition probability increases (sensor signal
decreases) with $\tau$ due to spin-relaxation  with a
characteristic time scale of 70$\pm$10 $\mu$s. A best-fit plane
has been subtracted from the data.}
\label{figurename}\end{center}\end{figure}

The singlet-triplet relaxation time can be determined by measuring
the time dependence of the charge sensing signal inside of the
pulse triangle. We extract $\tau_{ST}$ by measuring $G_{S2}$ as a
function of the pulse train period, $\tau$. $G_{S2}$ is measured
inside the pulse triangle ($V_R$,$V_L$ held  fixed at -403,-523.8
mV, respectively) and is plotted as a function of $\tau$ in Fig.\
4. In (1,1), $G_{S2}$$ \sim $20$\times$10$^{-3}$e$^2$/h, whereas
outside the pulse triangle in (0,2), $G_{S2}$$ \sim
$10$\times$10$^{-3}$e$^2$/h. For small $\tau$, $G_S$$ \sim
$15$\times$10$^{-3}$e$^2$/h in the pulse triangle. At long $\tau$,
$G_S$ approaches 10$\times$10$^{-3}$e$^2$/h in the pulse triangle,
which indicates complete transfer from the (1,1) to (0,2) charge
state. We fit these experimental data assuming exponential
singlet-triplet relaxation and find a best fit $\tau_{ST}$$
=$70$\pm$10 $\mu$s. The dependence of the singlet-triplet
relaxation time on detuning and magnetic field has been presented
in \cite{Johnson_Nature_2005}. A speed up of spin relaxation near
zero field is observed and is consistent with a hyperfine mediated
spin relaxation process.

\section{Coherent spin manipulation}
With an external magnetic field applied so that the $m_s=\pm1$
triplet states are split off, the $(1,1)_S$ and $(1,1)_{T0}$
states form a logical qubit. Due to the large singlet triplet
splitting in (0,2) we can easily initialize the system in
$(0,2)_S$. $(0,2)_S$ can transferred to $(1,1)_S$ by sweeping the
detuning adiabatically with respect to the interdot tunnel
coupling. The $(1,1)_S$ and $(0,2)_S$ states are hybridized at
zero detuning due to the interdot tunnel coupling. This
hybridization results in an exchange splitting between the
$(1,1)_S$ and $(1,1)_{T0}$ states, $j(\epsilon)$. For large
negative detunings, $j(\epsilon)\rightarrow0$. The exchange
$j(\epsilon)$ can be tuned on ns timescales by applying pulses to
the gates defining the double dot. Using this fast control of the
exchange energy we have recently implemented a spin SWAP operation
\cite{Petta_Science}.
\\
\\
We acknowledge useful discussions with Sankar Das Sarma,
Hans-Andreas Engel, Xuedong Hu, Daniel Loss, Emmanuel Rashba, and
Peter Zoller. Funding was provided through the ARO under
DAAD55-98-1-0270 and DAAD19-02-1-0070, the DARPA-QuIST program,
and the NSF under DMR-0072777, the Harvard Center for Nanoscale
Systems, and the Sloan and Packard Foundations.

\end{document}